\documentclass[3p,times,authoryear,12pt]{elsarticle}
\usepackage[T1]{fontenc}
\usepackage{graphicx}
\usepackage{epstopdf}
\usepackage{amssymb}
\usepackage{amsfonts,amsmath,bm,bbm}
\usepackage{mathtools}
\usepackage{enumerate}
\usepackage{indentfirst}
\usepackage[font=footnotesize]{caption}
\usepackage{floatrow}
\usepackage{enumitem}
\usepackage[table,xcdraw,dvipsnames]{xcolor}
\usepackage{url}
\usepackage{multirow}
\usepackage{natbib}
\usepackage[normalem]{ulem}
\useunder{\uline}{\ul}{}
\usepackage{setspace}

\usepackage{multicol}
\usepackage{enumitem}

\usepackage{tikz}
\usepackage{varwidth}
\usetikzlibrary{patterns,decorations.pathreplacing}

\newcommand{\sgn}{\text{sgn}}

\definecolor{dcyan}{rgb}{0.1,0.3,0.6}
\long\def\rev#1{{{#1}}}

\begin{document}

\journal{Electric Power Systems Research}

\begin{frontmatter}

\title{Variance Stabilizing Transformations for Electricity Price Forecasting in Periods of Increased Volatility}

\author[inst1]{Bartosz Uniejewski}

\affiliation[inst1]{organization={Department of Operations Research and Business Intelligence},
 addressline={Wroc{\l}aw University of Science and Technology}, 
 city={Wroc{\l}aw},
 postcode={50-370},
 country={Poland}}

% \cortext[cor1]{Corresponding author}

\begin{abstract}
Accurate day-ahead electricity price forecasts are critical for power system operation and market participation, yet growing renewable penetration and recent crises have caused unprecedented volatility that challenges standard models. This paper revisits variance stabilizing transformations (VSTs) as a preprocessing tool by introducing a novel parametrization of the asinh transformation, systematically analyzing parameter sensitivity and calibration window size, and explicitly testing performance under volatile market regimes. Using data from Germany, Spain, and France over 2015–2024 with two model classes (NARX and LEAR), we show that VSTs substantially reduce forecast errors, with gains of up to 14.6\% for LEAR and 8.7\% for NARX relative to untransformed benchmarks. The new parametrized asinh consistently outperforms its standard form, while rolling averaging across transformations delivers the most robust improvements, reducing errors by up to 17.7\%. Results demonstrate that VSTs are especially valuable in volatile regimes, making them a powerful tool for enhancing electricity price forecasting in today’s power markets.
\end{abstract}

\begin{keyword}
Electricity price forecasting, power market, Variance Stabilizing Transformation, adaptive model averaging, increased volatility
\end{keyword}

\end{frontmatter}

\section{Introduction and motivation}

Electricity price forecasting (EPF) is essential to modern power system operation \citep{mac:uni:wer:23,tan:etal:23}. Accurate day-ahead price forecasts support critical tasks such as unit commitment, reserve scheduling, demand bidding, and risk management for utilities and market participants alike \citep{zha:etal:22}. As electricity markets become liberalized and more interconnected, reliable forecasting becomes even more essential for ensuring informed system operation and cost-effective market participation \citep{mac:lip:uni:25}.

\rev{Beyond market participation and bidding strategies, electricity price forecasting also supports short-term operational decision making in power systems. As discussed by \cite{wer:14}, short-term price forecasts provide information relevant for operational planning under high volatility, where price spikes and scarcity events reflect underlying system stress. On the demand side, price forecasts are used to guide the operational scheduling of flexible assets such as energy storage systems, where charging and discharging decisions depend directly on expected price dynamics \citep{chi:zam:zar:pal:18}. Moreover, forecasting errors can lead to economically suboptimal operational schedules, particularly during volatile periods, highlighting the importance of forecast robustness beyond purely statistical accuracy \citep{zar:can:bha:10}.}

At the same time, forecasting electricity prices has become substantially more difficult \citep{lag:mar:des:wer:21}. The rapid expansion of renewable energy sources, together with regulatory changes and geopolitical shocks, has led to unprecedented price volatility \citep{gia:par:pel:16, mar:bel:ren:19}. Day-ahead prices now exhibit statistical features that are especially challenging for forecasting: sharp spikes, heavy-tailed distributions, strong seasonal cycles, and --increasingly since the mid-2010s -- frequent zero or even negative values. Such characteristics violate the assumptions underlying many classical time-series approaches and degrade their performance in exactly the conditions when accurate forecasts are most valuable.

A promising method for improving forecast robustness in electricity markets is the use of variance-stabilizing transformations (VSTs). These transformations are applied to price series prior to model estimation to reduce the influence of extreme values -- especially crucial given the frequent spikes in power prices \citep{cia:mun:zar:22}. By stabilizing volatility, VSTs improve the statistical properties of the data and enhance the reliability of forecasting models \citep{jan:tru:wer:wol:13}. Historically, the logarithmic transformation was the most common choice due to its simplicity. However, the widespread emergence of zero and negative day-ahead prices rendered the log transform infeasible, motivating the search for alternatives. Since the comprehensive evaluation by \citet{uni:wer:zie:18}, which confirmed that well-chosen VSTs can markedly enhance forecast accuracy they have become widely used in EPF literature \citep[for example][]{mar:bel:ren:19,kat:zie:21}. 

\citet{uni:wer:zie:18} systematically assessed multiple transformations across 12 electricity markets and two model classes, showing that well-chosen VSTs can improve forecast accuracy. However, that study adopted fixed parameter values for each transformation, relied on a single calibration window size, and was based on data from a relatively stable period (2010–2016), predating the recent surge in market volatility. Similar framework with fixed parameter values was adopted in many paper that compared different transformation function in EPF context \citep{shi:etal:21,cia:mun:zar:22}

\rev{\cite{jed:lag:mar:wer:22} review the evolution of electricity price forecasting methods, from classical time-series and regression models to regularized statistical learning approaches and, more recently, deep-learning and hybrid architectures. While early EPF relied primarily on autoregressive formulations \citep{wer:14}, regularization techniques such as LASSO enabled robust high-dimensional models including the LEAR framework \citep{zie:16:TPWRS} Although deep-learning models have shown promising results, large-scale empirical evidence indicates that well-designed statistical learning models remain highly competitive at the day-ahead horizon \citep{wer:14,lag:mar:des:wer:21,jed:lag:mar:wer:22} Since variance-stabilizing transformations operate at the preprocessing stage, they are applicable across different forecasting paradigms; however, their impact is expected to diminish as model nonlinearity increases, with highly nonlinear architectures such as deep neural networks potentially mitigating the influence of extreme observations intrinsically. Accordingly, this study focuses on model classes for which preprocessing plays a key role in estimation stability and forecast robustness.}

This paper revisits the problem of variance stabilization in EPF using a contemporary dataset spanning 2015–2024, a period that includes both the COVID-19 shock and the 2022 European energy crisis. Our contributions are threefold:
\begin{enumerate}
    \item Parameter sensitivity and calibration depth. We show that relying on fixed parameter settings, as done in earlier studies, is inadequate. Optimal parameters vary substantially across markets and calibration window sizes, and transformations perform inconsistently when the amount of historical data changes. This highlights the need for systematic parameter exploration.
    \item Novel transformation and adaptive schemes. We propose a new parametrization of the inverse hyperbolic sine (asinh) transformation, introducing a parameter related to slope at the origin. This parametrization consistently outperforms the standard form. In addition, we evaluate adaptive schemes for selecting or averaging across transformations and parameterizations. Among these, rolling averaging proves most effective, providing robust gains over fixed strategies.
    \item Volatility-regime analysis. We explicitly assess how the benefits of VSTs depend on market conditions by splitting the out-of-sample period into stable and volatile phases. Results demonstrate that transformations are particularly valuable during periods of heightened volatility, when reliable forecasts are most critical for power system operation.
\end{enumerate}
% By addressing methodological limitations of earlier studies and aligning the analysis with recent more volatile electricity markets, this paper provides new evidence on when and how VSTs enhance electricity price forecasting, and demonstrates their practical value as a tool for improving decision-making in liberalized power systems.

The rest of the paper is structured as follows. Section \ref{sec:data} describes the datasets used, including calibration window design. Section \ref{sec:methodology} outlines the methodological framework, beginning with the baseline forecasting models, followed by the description of variance-stabilizing transformations, the proposed selection and averaging schemes, and the evaluation measures. Section \ref{sec:Results} presents the empirical results, with emphasis on parameter sensitivity, transformations comparison, and temporal performance under volatile conditions. Section \ref{sec:conclusion} concludes with the main findings, limitations, and directions for future research.

\section{Data}
\label{sec:data}

To thoroughly evaluate the studied models, as done by \citet{gon:lin:23}, and in accordance with the best practices outlined by \citet{lag:mar:des:wer:21}, we utilize very long test sets from multiple markets. We use day-ahead electricity market data from three structurally and geographically diverse power systems: Germany (EPEX-DE), France (EPEX-FR), and Spain (OMIE). The dataset spans the period from 1.1.2015 to 31.12.2024, offering a consistent and extended basis for comparative analysis across forecasting models.

The German EPEX-DE market is one of the most widely studied markets for electricity price forecasting \citep{zie:wer:18, cia:mun:zar:22}, owing to its central position in Europe, high trading liquidity, and complex price dynamics, including frequent negative prices. The market is also characterized by a significant proportion of intermittent renewables; in 2024, the installed capacity of wind and solar power farms accounted for 58\% of total installed capacity. In addition to hourly day-ahead prices, we include day-ahead forecasts of electricity demand and renewable (wind and solar) generation

\begin{figure*}[tb!]
 \centering
 \includegraphics[width = 0.99\linewidth]{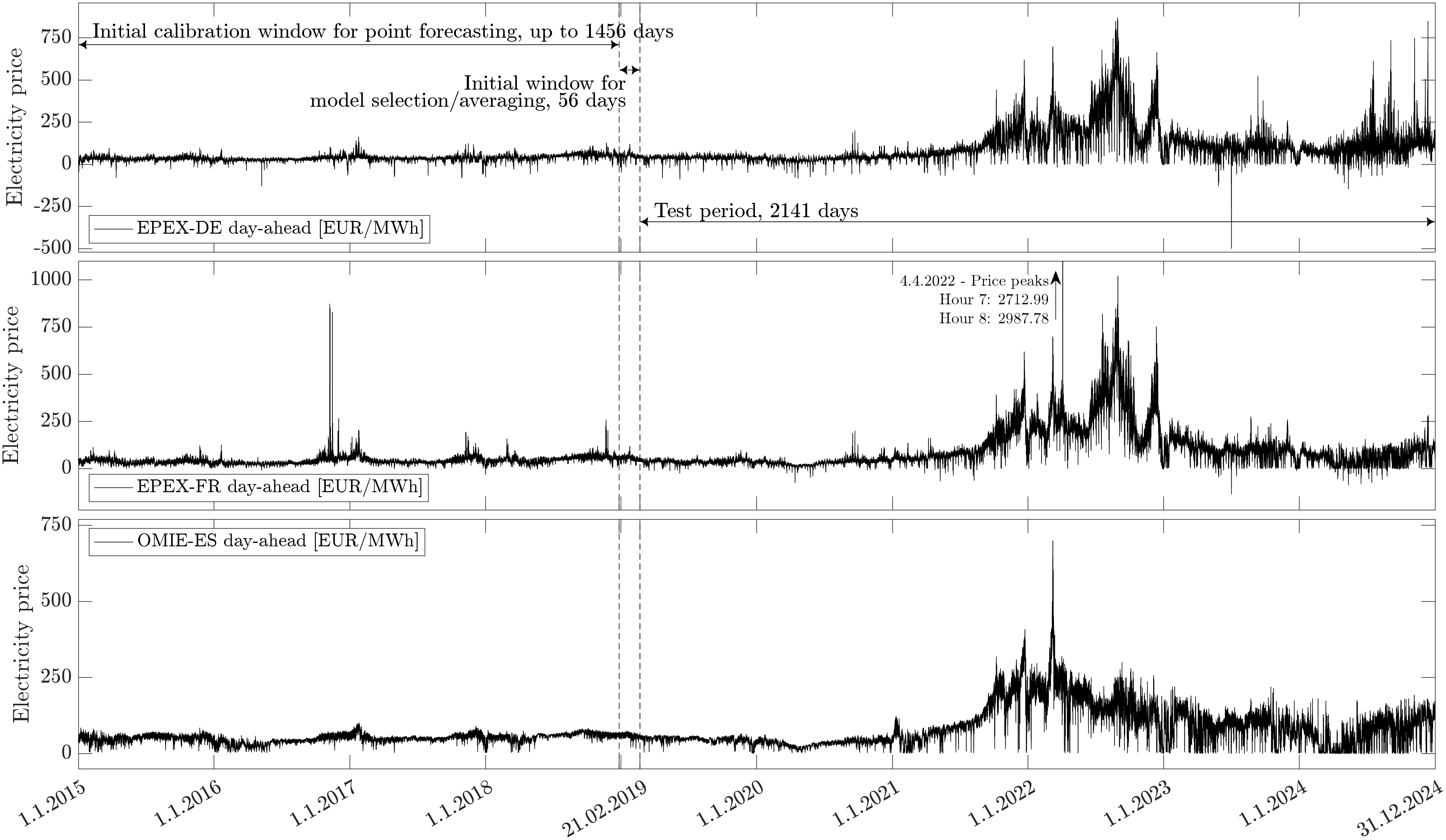}
 \caption{EPEX-DE (top), EPEX-FR (middle), OMIE-ES (bottom) hourly day-ahead prices for the period 1.1.2015-31.12.2024. The first vertical dashed line marks the end of the 1456-day calibration window for point forecasting models and the beginning of the 56-day period for selection/averaging methods. The second dashed line marks the beginning of the 2141-day out-of-sample test period.
 } 
 \label{fig:data}
\end{figure*}

The French EPEX is characterized by a unique generation mix dominated by nuclear power, but in recent years has also seen a gradual rise in renewable energy integration, particularly wind. It plays a key role in cross-border electricity flows within Western Europe \citep{sze:bak:23} and is increasingly exposed to price volatility driven by both internal and regional dynamics. As in the case of Germany, we use hourly day-ahead prices and incorporate TSO forecasts of electricity load and renewable generation to capture the growing impact of renewables on price formation.

Spanish OMIE market has become increasingly important due to its rapidly changing generation mix. Between 2015 and 2024, Spain’s solar power installed capacity share quadrupled — from around 5\% to over 20\%. Similarly, for Spain, we include hourly day-ahead prices, along with TSO forecasts of load and renewable generation. This allows for consistent modeling inputs across all three markets.

\rev{All the time series were obtained from a publicly available source, namely, the ENTSO-E Transparency Platform. (\url{transparency.entsoe.eu/})}\footnote{\rev{The day-ahead electricity prices were downloaded from from \url{transparency.entsoe.eu/market/energyPrices} for corresponding bidding zones (BZN) FR for France, ES for Spain and the DE-AT-LU bid zone until 30.9.2018 and DE-LU afterwards () for Germany. Both day-ahead total load forecasts (\url{transparency.entsoe.eu/market/energyPrices}) and day-ahead wind and solar generation forecasts (\url{transparency.entsoe.eu/generation/forecast/windAndSolar/}) were downloaded for contries (CTY) Germany, France and Spain.}} \rev{All data were downloaded in hourly or 15-minutes resolution and the latter one} have been processed to an hourly resolution. To ensure consistency, adjustments were made to account for missing data and changes to daylight saving time. Specifically, missing timestamps (e.g., during the spring shift) were imputed using the average of the preceding or following hour or day. Duplicate timestamps (e.g., during the fall shift) were replaced by their mean\footnote{\rev{The codes to preprocess the data can be provided from authors}}.

Additionally, we have enriched the dataset with commodity market indicators, recognizing the influence of global energy markets on electricity prices. These indicators include the closing prices of coal (API2), natural gas (TTF), crude oil (Brent), and carbon emission allowances (EUA), sourced from \url{Investing.com}\footnote{\rev{EUA: \url{www.investing.com/commodities/carbon-emissions} \\ Brent: \url{www.investing.com/commodities/brent-oil-historical-data} \\
Coal: \url{www.investing.com/commodities/newcastle-coal-futures-historical-data} \\
Gas: \url{www.investing.com/commodities/dutch-ttf-gas-c1-futures-historical-data}}}.

\subsection{Calibration window size}

Previous studies \citep{mar:ser:wer:18} have shown that the choice of calibration window size is critical to electricity price forecasting and that there is no universally optimal length. Although longer windows generally reduce estimator variance and are advantageous under stationarity, electricity markets are highly non-stationary due to structural shifts, such as increasing renewable energy source (RES) penetration, and exogenous shocks, such as the pandemic caused by the COVID-19 virus and the energy crisis triggered by the war in Ukraine. These changes may make long historical windows less relevant to current price dynamics. Conversely, shorter windows can adapt more quickly to recent patterns, though they may lack sufficient data for stable model estimation.

The goal of this study is not to identify the optimal window size, but rather to assess whether the performance of different VSTs depends on the sample size of the historical data used for model calibration. We considered five calibration windows ranging from approximately two months to four years: 56, 182, 364, 728, and 1,456 days.

\section{Methodology}
\label{sec:methodology}
	Point forecasts are generated using a rolling calibration window scheme with five different window lengths. For each window size, the forecasting models described in Section~\ref{ssec:models} are estimated using historical data preceding the forecast day. For example, to produce the forecast for 27.12.2018 (it is the first forecasted day), we are using data from 01.01.2015 (for 1456-day window, but only from 01.11.2018 for 56-day window) to 26.12.2018. The window is then shifted forward by one day, and the next forecast is generated. This rolling procedure is repeated until forecasts for all 24 hours of 31.12.2024 are obtained, resulting in 2197 forecast days.

 After computing the point forecasts, a separate 56-day window is used to calibrate the selection or averaging method. This step identifies the best-performing transformation (or ensemble of transformations) based on recent forecast errors to determine how to use them for final prediction.
	
\subsection{Base Models}

The forecasting models considered in this study are used as established benchmarks \citep{mac:lip:uni:25} to evaluate the impact of variance-stabilizing transformations rather than to introduce new modeling architectures. The comparison between NARX and LEAR is intended to assess the robustness of the proposed transformation and selection schemes across models with different structural characteristics. A detailed analysis of model complexity or interpretability is outside the scope of this paper.

\label{ssec:models}

\subsubsection{NARX model}

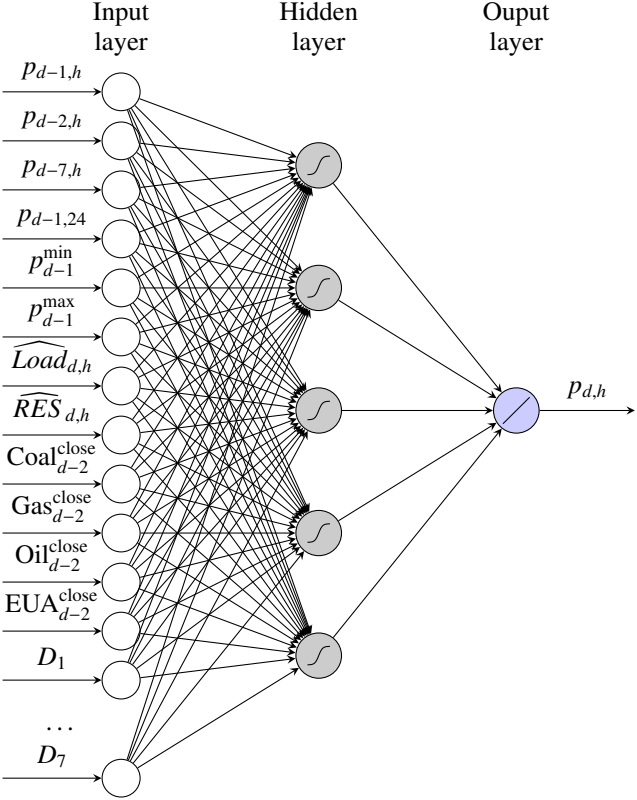
\begin{figure}[b!]
	\tikzset{%
		input neuron/.style={
			circle,
			draw,
			% fill=blue!50,
			minimum size=.5cm
		},
		every neuron/.style={
		circle,
		draw,
		% fill=blue!50,
		minimum size=.6cm
		},
		neuron missing/.style={
			draw=none, 
			scale=1.2,
			text height=.25cm,
			execute at begin node=\color{black}$\vdots$
		},
		sigmoid/.style={path picture= {
				\begin{scope}[x=.7pt,y=7pt]
					\draw plot[domain=-6:6] (\x,{1/(1 + exp(-\x))-0.5});
				\end{scope}
			}
		},
		linear/.style={path picture= {
			\begin{scope}[x=5pt,y=5pt]
				\draw plot[domain=-1:1] (\x,\x);
			\end{scope}
		}
	},
	}

	\centering
	\begin{tikzpicture}[x=1.5cm, y=.7cm, >=stealth]
	
	\foreach \m/\l [count=\y] in {1,2,3,4,5,6,7,8,9,10,11,12,13}
	\node [input neuron/.try, neuron \m/.try] (input-\m) at (0,2.5-\y) {};

 \foreach \m/\l [count=\y] in {14}
	\node [input neuron/.try, neuron \m/.try] (input-\m) at (0,-12.5) {};
	
	\foreach \m [count=\y] in {1,2,3,4,5}
	\node [every neuron/.try, neuron \m/.try, fill= black!20, sigmoid ] (hidden-\m) at (2.,2.5-\y*2.5) {};
	
	\foreach \m [count=\y] in {1}
	\node [every neuron/.try, neuron \m/.try, fill= blue!20, linear ] (output-\m) at (4,-5) {};
	
	\foreach \l [count=\i] in {$p_{d-1,h}$,$p_{d-2,h}$,$p_{d-7,h}$,$p_{d-1,24}$,$p_{d-1}^{\min}$,$p_{d-1}^{\max}$,$\widehat{Load}_{d,h}$, $\widehat{RES}_{d,h}$, $\text{Coal}_{d-2}^{\text{close}}$, $\text{Gas}_{d-2}^{\text{close}}$, $\text{Oil}_{d-2}^{\text{close}}$, $\text{EUA}_{d-2}^{\text{close}}$, $D_1$,$D_7$}
	\draw [<-] (input-\i) -- ++(-1.2,0)
	node [above, midway] {\footnotesize\l};

 \node[] at (-0.6,-11.5) {\footnotesize \ldots};
 
	\foreach \l [count=\i] in {1,2,3,4,5}
	\node [above] at (hidden-\i.north) {};
	
	\foreach \l [count=\i] in {1}
	\draw [->] (output-\i) -- ++(1.2,0)
	node [above, midway] {\footnotesize{$p_{d,h}$}};
	
	\foreach \i in {1,...,14}
	\foreach \j in {1,...,5}
	\draw [->] (input-\i) -- (hidden-\j);
	
	\foreach \i in {1,...,5}
	\foreach \j in {1}
	\draw [->] (hidden-\i) -- (output-\j);
	
	\foreach \l [count=\x from 0] in {Input, Hidden, Ouput}
	\node [align=center, above] at (\x*2,2) {\footnotesize\l \\[-3pt] \footnotesize{layer}};
	
	\end{tikzpicture}	
	\caption{\rev{Visualization of the NARX network with five hidden neurons with hyperbolic tangent activation functions and one linear output neuron. Source: \citet{lip:uni:25}}}
	\label{fig:NARX}
\end{figure}

The first base model used in this study is a feedforward neural network within the NARX (nonlinear autoregression with exogenous variables) framework. This model is implemented in a series-parallel architecture \citep{xie:tan:lia:2009}. The model setup, including the selection of inputs and the network architecture, was adopted directly from \citet{lip:uni:25}, where it served as the baseline for point forecasting. The network consists of a single hidden layer with five neurons \rev{and a subsequent linear layer producing a single-valued output (see figure \ref{fig:NARX}) The state of the i-th neuron in the hidden layer can be expressed as:
$$
h_i = f(\sum_{j=1}^n x_i w^{(h)}_{ji} + b^{(h)}_i),
$$
where $x_j$ is the j-th regressor and $n$ is the number of regressors. $f(\cdot)$ is a nonlinear activation function, in this study selected to be a hyperbolic tangent function. The final output layer is a linear combination of the hidden-layer neuron states:
$$
\hat{y} = \sum_{i=1}^5 h_i w^{(o)}_{i} + b^{(o)}.
$$
Model parameters are estimated in Matlab using the \texttt{trainlm} function, which implements the Levenberg--Marquardt optimization algorithm \citep{hag:men:1994}. To mitigate overfitting, early stopping is applied: training is halted when the validation error, computed using a 9:1 training-to-validation split, does not improve for 10 consecutive epochs.}

The inputs include lagged electricity prices, daily price extremes, the most recent price at midnight, day-ahead forecasts of load and renewable energy source (RES) generation, commodity prices (coal, gas, oil, and EUAs), and weekday dummies — consistent with the best practices established in earlier studies \citep{bil:gia:del:rav:23,che:uni:wer:25}. A separate model is trained for each hour using a rolling window approach. To reduce estimation variance, an ensemble of ten independently trained NARX models is used for each forecast, and the final prediction is taken as the average across ensemble members.

\subsubsection{LEAR model}

The second forecasting approach used in this study is a high-dimensional linear model, often referred to in the EPF literature as the LEAR (LASSO Estimated AutoRegressive ) model. Originally introduced by \citet{uni:now:wer:16} and \citet{zie:16:TPWRS} the LEAR framework captures complex temporal and cross-hour dependencies by including a large number of predictors. Estimation is handled via LASSO regression \citep{tib:96} to control for overfitting. \rev{Prior to estimation, all explanatory variables are standardized to have a mean of zero and a variance of one, while the dependent variable remains on its original scale. The LASSO penalty parameter is selected from a grid of 100 values via 7-fold cross-validation within each calibration window.}

The model includes lagged prices for all 24 hours on days $d-1$ and $d-7$, daily price extremes ($p^{\min}_{d-1}$ and $p^{\max}_{d-1}$), and day-ahead forecasts of load, and RES generation for all 24 hours on day $d$. It also incorporates the most recent closing prices (from day $d-2$) of key commodities -- coal, natural gas, crude oil, and EU carbon emission allowances -- as well as a full set of seven weekday dummies and \rev{non-penalized} intercept. In total, the model includes over 150 regressors, enabling it to capture rich dependencies across time and variables. As with the NARX model, forecasts are generated independently for each hour of the day using a rolling window estimation approach.
\begin{align}
 p_{d,h} & = \beta_0 + \sum_{i=1}^{24} \left( \beta_i p_{d-1,i} + \beta_{24+i} p_{d-7,i} \right) + \beta_{49} p_{d-1}^{\min} + \beta_{50} p_{d-1}^{\max} \nonumber\\ 
 & + \sum_{i=1}^{24} \beta_{50+i} \widehat{\text{Load}}_{d,i} + \beta_{74+i} \widehat{\text{Load}}_{d-1,i}\nonumber\\
 & + \sum_{i=1}^{24} \beta_{98+i} \widehat{\text{RES}}_{d,i} +\beta_{122+i} \widehat{\text{RES}}_{d-1,i} 
 \nonumber\\
 & + \beta_{147} \text{Coal}_{d-2}^{\text{close}}+ \beta_{148} \text{Gas}_{d-2}^{\text{close}} + \beta_{149} \text{Oil}_{d-2}^{\text{close}} \nonumber\\ &+ \beta_{150} \text{EUA}_{d-2}^{\text{close}}
 + \sum_{i=1}^7 \beta_{150+i} D_i + \varepsilon_{d,h}.
 \label{eq:HLM}
\end{align}

\subsection{Variance Stabilizing Transformations}

Variable stabilizing transformations (VSTs) aim to reduce the impact of extreme observations on model estimates, which can cause underperformance of the employed predictive algorithms. VSTs improve the statistical properties of the data, allowing for more stable and accurate point forecasts.

Before applying any variance stabilizing transformation, the electricity price series is standardized using the median–median absolute deviation (MAD) pair, following the recommendation of \citet{uni:wer:zie:18}. This robust alternative to mean–standard deviation scaling is well-suited to electricity price data, which often exhibits outliers and price spikes. For each calibration window, the data are centered by subtracting the median and scaled by the MAD (median absolute deviation), which is defined as the median of the absolute deviations from the median. This standardization step improves the numerical stability by ensuring that all transformations operate on a centered, scale-adjusted inputs.

This study focuses on four transformations: the inverse hyperbolic sine (asinh) with a tunable slope parameter; the Box-Cox transformation; the mirror log (mlog) transformation; and the probability integral transform (PIT) with a t-distribution. These choices are based on two criteria: (i) strong empirical performance in the benchmark comparison conducted by \citet{uni:wer:zie:18, cia:mun:zar:22} and (ii) continued relevance and adoption in the EPF literature.

\paragraph{Parametrized Inverse Hyperbolic Sine (asinh)}
The asinh transformation was first introduced to the EPF literature by \citet{sch:11} as a way to stabilize variance without losing interpretability for small values. Since then it has been employed in multiple studies in combination with variety of models \citep{shi:etal:21,gro:kac:kru:23,jan:25}. Its appeal lies in its ability to behave linearly around the origin while still compressing large positive or negative values. In this work we extend the classical form by introducing a tunable slope parameter $c>0$:
\begin{equation}\label{eqn:asinh}
Y_{d,h} = \sgn(p_{d,h}) \left[ \text{asinh}\left(|p_{d,h}|+\sqrt{\frac{1}{c^2}-1}\right) - \text{asinh}\left(\sqrt{\frac{1}{c^2}-1}\right) \right],
\end{equation}

where $c > 0$ is a tuning parameter that controls the slope at the origin. This parametrization, introduced here for the first time, allows for adaptation to different volatility regimes, providing greater flexibility than the standard asinh. We consider $c$ values in the range $\{0.1, \dots, 1\}$. Note that, up to this point, the unparametrized asinh transformation is equivalent to the proposed version with $c=1$.

\paragraph{Box-Cox Transformation}
The Box–Cox family is one of the most widely used variance-stabilizing transformations in time series analysis \citep{hyn:ath:13}. In its standard form it is not defined for non-positive values, which limits its applicability to electricity prices in markets with zero or negative outcomes. To address this limitation, we employ a robust variant proposed by \citet{sak:92}, denoted here as \textbf{boxcox}$(\lambda)$ and defined as:
\begin{eqnarray}
Y_{d,h} = \sgn(p_{d,h}) \left\{ \begin{array}{ll}
\frac{\left( |p_{d,h}|+1\right) ^\lambda -1}{\lambda} & \textrm{for $\lambda>0$},\\
\log(|p_{d,h}|+1) & \textrm{for $\lambda=0$},\\
\end{array} \right. 
\end{eqnarray}

where $\lambda$ is a shape parameter controlling the degree of variance stabilization. In the EPF literature, Box–Cox has been widely applied, often with $\lambda=0.5$ as a default choice \citep{cia:mun:zar:22,lip:uni:25}. Here we broaden the analysis by considering $\lambda \in \{0.0,0.1,\dots,1.0\}$ to examine how the optimal setting depends on market conditions and calibration window length.

\paragraph{Mirror Log (mlog)}
The mirror logarithm was introduced in the EPF context by \citet{uni:wer:zie:18} as a robust alternative to the standard logarithm. It is a straightforward generalization of the log-transform with a mirror image of the logarithm for negative values. Subsequent studies have employed mlog in both point and probabilistic forecasting models \citep{kat:zie:18, leh:sch:her:22}. The transformation is defined as:
\begin{equation}
	Y_{d,h} = \sgn(p_{d,h})\left[\log\left(|p_{d,h}|+\tfrac{1}{c}\right) + \log(c)\right].
\end{equation}

The mlog transformation is constructed to have a slope of c at the origin. Therefore, mlog(1) is equivalent to boxcox(0). Based on the limited testing \citet{uni:wer:zie:18} suggest to take $c=\frac{1}{3}$, here the method is evaluated for $c$ over the grid $\{0.1, \dots, 1\}$.

\paragraph{T-PIT: Probability Integral Transform with $t$-Distribution.}
The PIT framework is grounded in transforming data via its empirical cumulative distribution function, followed by mapping through the inverse of a reference distribution. In EPF, the normal distribution has been the standard choice, with successful applications in both linear and nonlinear forecasting models \citep{shi:etal:21,cia:mar:nas:23}. However, normal-based PIT may not sufficiently accommodate heavy-tailed behavior in volatile markets. To address this, we adopt the inverse $t$-distribution with $\nu$ degrees of freedom:
\begin{equation}\label{eqn:N-PIT}
Y_{d,h} = G^{-1}\left(\text{PIT}(P_{d,h})\right) = G^{-1}\left(\hat{F}_{P_{d,h}}(P_{d,h})\right),
\end{equation}
where $G^{-1}$ is the inverse of $t$-distribution. We evaluate the method for the grid of $\nu$ in the range $3^2, 4^2, \ldots 7^2$. Note that in the EPF literature, the PIT transformation is usually associated with a normal distribution rather than a $t$-distribution. However, a $t$-distribution with 128 degrees of freedom can be used to approximate a normal distribution.

\subsection{Forecast Selection and Combination Methods}
\label{ssec:selection_combination}

Once point forecasts have been generated using multiple parameter configurations of a single variance stabilizing transformation (VST), or across several distinct VSTs, the next step is to determine which of these forecasts should be used. Relying on a single transformation or fixed parameter may lead to suboptimal results. To address this, we propose four systematic approaches to post-process the pool of forecasts. We divide those methods into two categories: \textit{selection-based methods}, which choose a single forecast from the set, and \textit{averaging-based methods}, which combine multiple forecasts. In all cases. the selection of forecasts is based on mean absolute error (MAE) within a 56-day rolling or fixed window. The procedures are applied independently for each hour of a day.

\rev{It is important to note that the proposed procedures do not involve direct optimization or tuning of VST parameters within the forecasting models. 
Instead, multiple parameter configurations are considered in parallel, and the final forecast is selected or combined ex post based on recent forecasting performance, which constitutes the main contribution of this study.}

\subsubsection{Selection-Based Methods}

The selection-based methods aim to identify the single most accurate forecast from a set of individual predictions. In the \textit{static selection} approach, denoted by $\mathrm{SEL}_{\text{fix}}$, the best-performing transformation (or parameter configuration) is identified based on its MAE over the 56-days initial selection window (see Figure \ref{fig:data}). This selection is then used unchanged for the entire out-of-sample period. 
%For example, $\mathrm{SEL}_{\text{fix}}(\text{asinh})$ refers to selecting the best parameter configuration of the asinh transformation once at the beginning of the period and using the same transformation parameter throughout.

In contrast, the \textit{rolling selection} method, denoted by $\mathrm{SEL}_{\text{roll}}$, allows for temporal adaptation. Instead of relying on a single static decision, the best-performing transformation is re-evaluated each day using the most recent 56 days. This enables the forecast method to respond to structural changes or evolving market conditions. %When applied to a single transformation, we use notation such as $\mathrm{SEL}_{\text{roll}}(\text{mlog})$, indicating that the best parameter setting of mlog is reselected daily.

\subsubsection{Averaging-Based Methods}

Instead of choosing a single forecast, averaging-based methods combine multiple forecasts to increase robustness and reduce sensitivity to parameter or transformation mis-specification. In the \textit{static averaging} method, denoted $\mathrm{AVG}_{\text{fix}}$, we evaluate all combinations of one, two, or three forecasts based on their performance over the initial 56-days averaging window. The best-performing combination -- determined by the lowest MAE -- is selected and used for the remainder of the forecasting period. %When applied to a single transformation, this is written as $\mathrm{AVG}_{\text{fix}}(\text{tpit})$, for instance.

The \textit{rolling averaging} method, denoted $\mathrm{AVG}_{\text{roll}}$, updates the selected combination hourly using a rolling 56-day calibration window. For each hour, the combination of forecasts (up to three) with the best recent performance is identified and used. This allows the ensemble to adapt to changing forecast dynamics and maintain performance stability. 
%As with the other methods, when applied within a single transformation, we use notation such as $\mathrm{AVG}_{\text{roll}}(\text{boxcox})$.

The key distinction between the fixed and rolling variants in both selection and averaging families lies in their ability to adapt over time. While the fixed methods offer simplicity and computational efficiency, the rolling variants provide flexibility and resilience to market non-stationarities.

Each method can be applied either to a set of forecasts derived from \textit{different parameter values of a single VST} or to forecasts obtained from \textit{multiple VSTs}. In the former case, we indicate the specific transformation in parentheses, e.g., $\mathrm{AVG}_{\text{roll}}(\text{boxcox})$, while methods applied across transformations are denoted without such specification, e.g., $\mathrm{AVG}_{\text{roll}}$.

\subsection{Evaluation measures}

To evaluate the accuracy of point forecasts, we use the Mean Absolute Error (MAE), one of the most widely adopted performance metrics in electricity price forecasting. MAE measures the average absolute difference between the predicted and actual values, offering a straightforward and interpretable indication of forecast accuracy. The measure for a single day is defined by the following equation:

% Linear measures like mean absolute error assess the average forecasting error but focus on median forecasts, not mean predictions. This study uses the root mean square error (RMSE), defined by the following equation for a single day:

\begin{equation}
\text{MAE}_d = \sqrt{\frac{1}{24}\sum_{h=1}^{24} |\hat{\varepsilon}_{d,h}|}.
\label{eqn:MAEd}
\end{equation}

where $\hat{\varepsilon}_{d,h} = {p_{d,h} -\hat{p}_{d,h}}$ is the forecasting error for day $d$ and hour $h$. The aggregated measure for the whole out-of-sample period of $D$ days is defined as follows: $
\text{MAE} = \sqrt{\frac{1}{D}\sum_{d=1}^{D} (\text{MAE}_d)^2}$. \rev{All MAE values are expressed in EUR/MWh.}

In addition, we perform the conditional predictive ability \citep[CPA;][]{gia:whi:06} test to formally evaluate the performance of the considered models, by pairwise evaluation. For two selected models A and B, we test the null hypothesis $H_0:\boldsymbol{\phi}=0$ in following regression:
\begin{equation}\label{eqn:GW}
\Delta_d^{\text{A}, \text{B}}= \phi_0 + \phi_1 \Delta_{d-1}^{\text{A}, \text{B}} + \epsilon_d^{\text{A}, \text{B}}, 
\end{equation}
where $\Delta_{d}^{\text{A}, \text{B}} = \text{MAE}_{d}^{\text{A}} - \text{MAE}_{d}^{\text{B}}$, is the loss differential series and $\text{MAE}_{d}^i$ is the prediction errors of model $i$ for day $d$.

\rev{The aim of performing the CPA test is to statistically assess whether differences in forecast accuracy between two competing models are significant over time. In particular, it tests the null hypothesis of equal predictive performance by accounting for temporal dependence and heteroskedasticity in the loss differential series. A small p-value indicates that the null hypothesis can be rejected, implying that one model exhibits statistically different (superior or inferior) predictive performance relative to the other. The CPA test is applied pairwise to compare selected forecasting configurations and is used to support relative model assessment rather than as a simultaneous multiple-hypothesis testing procedure.}

\rev{In this paper, we focus on accuracy measures aggregated for all 24 hours. However an hour-by-hour accuracy analysis was also performed. Due to space limitations and the large number of models and configurations considered, these results are not included in the manuscript but are available from the author upon request.}

\section{Results}
\label{sec:Results}

This section presents the empirical results. We begin by examining the effect of parameterization on the performance of variance-stabilizing transformations (VSTs). Next, we compare individual transformations with adaptive selection and averaging schemes, before analyzing how these gains evolve under different market conditions.

\subsection{Impact of VST parameterization}
\begin{figure*}[b!]
 \includegraphics[width = 0.99\linewidth]{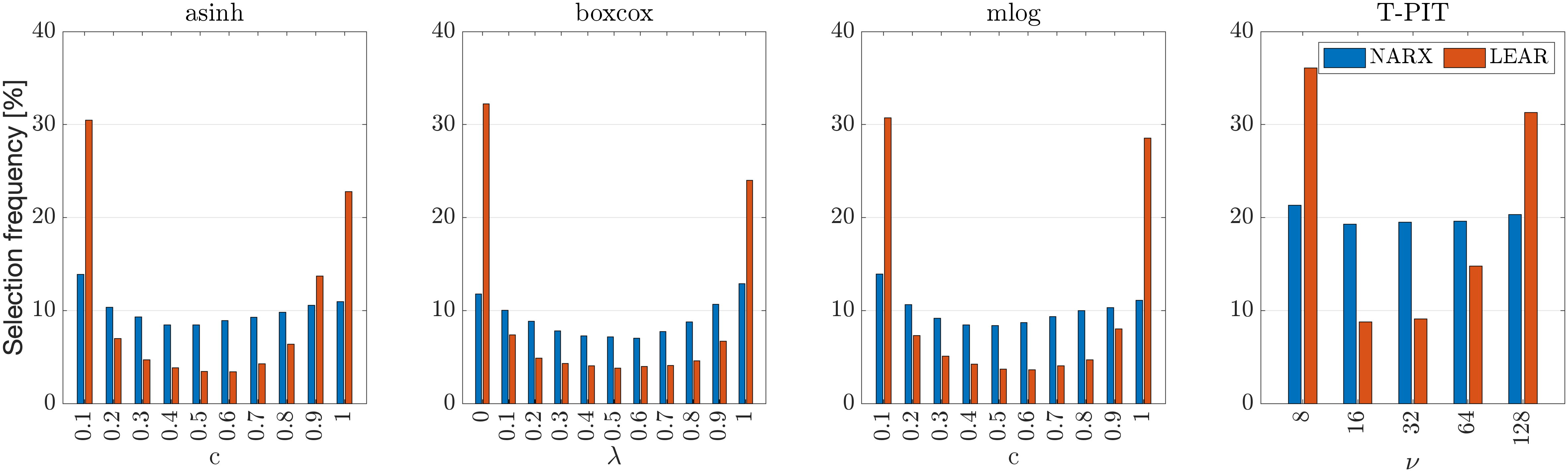}
 \caption{Empirical distribution of MAE-optimal parameter values for each variance stabilizing transformation (VST). Results are shown for the German EPEX market using the longest calibration window of 1,456 days, with separate distributions for the NARX and LEAR models. Optimal parameter is selected separately for each of 51,384 hours in out-of-sample period (2,141 days $\times$ 24 hours).
 Analogous plots for the Spanish and French markets and for other calibration window sizes show similar patterns and are therefore omitted for brevity.}
\label{fig:histogram}
\end{figure*}

Figure~\ref{fig:histogram} presents the empirical distribution of parameter values that yielded the lowest MAE across the four considered VSTs for the EPEX-DE market, based on the longest calibration window of 1456 days. Several observations can be made. First, no single parameter value consistently dominates, with some clustering at boundary values. Second, the two forecasting models exhibit different sensitivities. 
The LEAR model tends to favor more extreme parameterizations, whereas the NARX model shows a more balanced distribution, indicating lower dependence on precise parameter tuning. Finally, the wide dispersion of optimal parameter values underscores the fact that the relative performance of VSTs is highly context-dependent and cannot be reduced to a single universal setting. Note that the figure is provided only for the German EPEX market, as it aims to illustrate the motivation. Analogous plots for the Spanish and French markets, as well as for shorter calibration windows, reveal qualitatively similar behavior and do not change the conclusions. 

This result highlights the risk of relying on a single parameter choice. It also provides a strong rationale for the selection and averaging strategies which are designed to improve robustness across markets, models, and calibration window sizes.

\begin{table*}[htb!]
\caption{Mean absolute error (MAE) of forecasts for each variance stabilizing transformation (VST) and each market. 
Rows report results for: (i) \textit{CB} (crystal ball benchmark, using the ex-post optimal parameter for each hour), 
(ii) \textit{Literature} (fixed parameters used in prior studies), and (iii) four proposed selection and averaging schemes. The best result in each column, excluding the crystal ball benchmark, is highlighted in bold. }
\label{tab:mae_results}

\scalebox{0.75}{
\begin{tabular}{cr|ccc|ccc|ccc|ccc}
\multicolumn{1}{l}{} & & \multicolumn{3}{c|}{asinh} & \multicolumn{3}{c|}{boxcox} & \multicolumn{3}{c|}{mlog} & \multicolumn{3}{c}{T-PIT} \\
\hline
\hline
\multicolumn{1}{l}{} & & DE & ES & FR & DE & ES & FR & DE & ES & FR & DE & ES & FR \\
\multirow{6}{*}{\rotatebox{90}{NARX}} & CB & 5.87* & 4.69* & 1.21* & 6.21* & 4.29* & 2.02* & 4.90* & 4.43* & 2.54* & 10.54* & 4.39* & 2.63* \\
 & \textit{Literature} & 15.25 & 11.00 & 14.40 & 14.32 & 10.60 & 13.87 & 14.50 & 10.63 & 13.93 & 15.05 & 11.03 & 14.74 \\
 & $\mathrm{SL}_{\text{fix}}$ & 14.38* & 10.62* & 14.05* & 14.27* & 10.58 & 14.03 & 14.36* & 10.63 & 14.17 & 15.02 & 11.03 & 14.74 \\
 & $\mathrm{SL}_{\text{roll}}$ & 14.35* & 10.66* & 14.01* & 14.18 & 10.60 & 13.90 & 14.32* & 10.64 & 13.95 & 14.98 & 11.06 & 14.74 \\
 & $\mathrm{AVG}_{\text{fix}}$ & 15.06* & 11.00 & 13.95* & 14.54 & 10.51* & 13.78* & 14.36* & 10.70 & \textbf{13.82}* & 14.88* & 11.02 & 14.74 \\
 & $\mathrm{AVG}_{\text{roll}}$ & \textbf{14.28}* & \textbf{10.56}* & \textbf{13.87}* & \textbf{14.08}* & \textbf{10.50}* & \textbf{13.76}* & \textbf{14.25}* & \textbf{10.53}* & 13.85 & \textbf{14.87}* & \textbf{10.99} & \textbf{14.68}* \\
\hline
\hline
\multirow{6}{*}{\rotatebox{90}{LEAR}} & CB & 2.65* & 2.64* & 2.35* & 3.62* & 2.80* & 2.11* & 3.08* & 2.73* & 2.42* & 4.53* & 2.48* & 3.02* \\
 & \textit{Literature} & 15.57 & 11.54 & 14.32 & 14.44 & 10.16 & 13.47 & 14.31 & 10.20 & 13.41 & 16.21 & 11.53 & 15.20 \\
 & $\mathrm{SL}_{\text{fix}}$ & 14.61* & 10.09* & 13.70* & 15.02 & 10.15 & 14.39 & 14.60 & 10.08* & 13.70 & 16.17* & 11.53 & 15.60 \\
 & $\mathrm{SL}_{\text{roll}}$ & 14.02* & 10.06* & 13.20* & 13.94* & 10.01* & 13.18* & 13.98* & 10.07* & 13.20* & 15.92* & 11.47 & 15.08* \\
 & $\mathrm{AVG}_{\text{fix}}$ & 15.12* & 10.10* & 14.32 & 15.02 & 10.19 & 13.49 & 14.83 & 10.08* & 13.31* & 16.17* & 11.78 & 15.20 \\
 & $\mathrm{AVG}_{\text{roll}}$ & \textbf{13.97}* & \textbf{10.06}* & \textbf{13.18}* & \textbf{13.89}* & \textbf{9.99}* & \textbf{13.14}* & \textbf{13.92}* & \textbf{10.06}* & \textbf{13.17}* & \textbf{15.91}* & \textbf{11.47} & \textbf{15.08}*
\end{tabular}
}
\\[3pt]
Note: * indicate significant difference at the 5\% level of the \cite{gia:whi:06} test with respect to \textit{Literature}

\end{table*}

To address this issue, in Table~\ref{tab:mae_results} we report the mean absolute error (MAE) of forecasts obtained with different variance stabilizing transformations (VSTs) across the three markets, using the longest calibration window of 1456 days. The first row presents the crystal ball (CB) benchmark, which assumes that the optimal parameter is selected ex post for each hour. While unrealistic in practice, this benchmark highlights the potential gains achievable through parameter optimization, as it reduce the error by up to 90\% compared to model with fixed parameter utilized in prior studies. 

Moreover the results presented in Table~\ref{tab:mae_results} indicate a clear pattern, that emerges when comparing fixed versus rolling approaches: rolling schemes systematically outperform their fixed counterparts. Across models and transformations, MAE reductions of up to 5\% are typical when moving from static to rolling calibration, demonstrating the importance of updating parameter choices in line with recent market conditions. 

Among the proposed methods, the rolling averaging scheme ($\mathrm{AVG}_{\text{roll}}$) consistently delivers the strongest performance across markets and transformations. For example, with the asinh transformation under LEAR, MAE decreases by $-10.3\%$ in EPEX-DE, $-12.8\%$ in OMIE-ES, and $-8.0\%$ in EPEX-FR compared to the literature baseline with fixed parameters. For boxcox, mlog and T-PIT, the improvements are smaller (around $1$--$4\%$), but remain consistent. The conclusion is supported by the result of the \citet{gia:whi:06} test at the 5\% level. In nearly all cases, $\mathrm{AVG}_{\text{roll}}$ is significantly better than the literature benchmark, while others averaging or selection models -- especially with fixed schemes -- are less often significantly better.

\subsection{Transformation selection and averaging}

\setlength\tabcolsep{3.5pt}

\begin{table*}[htb!]
\caption{Mean absolute errors (MAE) for individual variance-stabilizing transformations (upper subsections) and proposed selection/averaging schemes (lower subsections) across five calibration window lengths. The best single transformation in each column is underlined, while the overall best result, including combination schemes, is shown in bold.}
\label{tab:results}
\scalebox{0.7}{
\begin{tabular}{cr|ccccc|ccccc|ccccc}
\multicolumn{1}{l}{} & & \multicolumn{5}{c|}{EPEX-DE} & \multicolumn{5}{c|}{OMIE-ES} & \multicolumn{5}{c}{EPEX-FR} \\ 
\multicolumn{1}{l}{} & & \multicolumn{1}{c}{56} & \multicolumn{1}{c}{182} & \multicolumn{1}{c}{364} & \multicolumn{1}{c}{728} & \multicolumn{1}{c|}{1456} & \multicolumn{1}{c}{56} & \multicolumn{1}{c}{182} & \multicolumn{1}{c}{364} & \multicolumn{1}{c}{728} & \multicolumn{1}{c|}{1456} & \multicolumn{1}{c}{56} & \multicolumn{1}{c}{182} & \multicolumn{1}{c}{364} & \multicolumn{1}{c}{728} & \multicolumn{1}{c}{1456} \\
\hline
\hline

\multirow{9}{*}{\rotatebox{90}{NARX}} & id & 17.46 & 15.08 & 14.21 & 14.47 & 14.40 & 12.29 & 11.14 & 10.74 & 10.82 & 10.64 & 16.89 & 14.84 & 14.40 & 14.33 & 14.11 \\
 & asinh & {\ul 15.94} & 14.34 & 13.83 & 14.15 & 14.28 & 11.46 & {\ul 10.85} & 10.67 & 10.73 & 10.56 & 15.57 & {\ul 14.27} & {\ul 14.05} & 14.05 & 13.87 \\
 & boxcox & 16.33 & 14.36 & 13.89 & 14.07 & {\ul 14.08} & 11.61 & 10.90 & 10.65 & 10.69 & {\ul 10.50} & 15.93 & 14.35 & 14.10 & {\ul 14.04} & {\ul 13.76} \\
 & mlog & 15.99 & {\ul 14.32} & 13.91 & 14.11 & 14.25 & {\ul 11.44} & 10.85 & 10.71 & 10.75 & 10.53 & {\ul 15.57} & 14.31 & 14.09 & 14.09 & 13.85 \\
 & T-PIT & 16.77 & 14.44 & {\ul 13.83} & {\ul 14.01} & 14.87 & 11.85 & 10.87 & {\ul 10.58} & {\ul 10.67} & 10.99 & 16.26 & 14.45 & 14.09 & 14.07 & 14.68 \\[2pt]
 \cline{2-17}
 & $\mathrm{SL}_{\text{fix}}$ & 16.33 & 14.36 & 13.91 & 14.11 & 14.25 & 11.46 & 10.87 & 10.71 & 10.73 & 10.56 & 16.26 & 14.45 & 14.09 & 14.09 & 13.87 \\
 & $\mathrm{SL}_{\text{roll}}$ & 16.12 & 14.36 & 13.80 & 13.88* & 14.07* & 11.53 & 10.82* & 10.53 & 10.51* & 10.40* & 15.68 & 14.30 & 13.99 & 13.83* & 13.71* \\
 & $\mathrm{AVG}_{\text{fix}}$ & 16.14 & 14.36 & 13.76 & 13.83* & 14.17 & 11.46 & 10.69* & 10.59 & 10.66 & 10.56 & 15.52 & \textbf{14.06}* & 14.09 & 13.94* & 13.87 \\
 & $\mathrm{AVG}_{\text{roll}}$ & \textbf{15.88} & \textbf{14.20}* & \textbf{13.69}* & \textbf{13.76}* & \textbf{13.98}* & \textbf{11.34}* & \textbf{10.68}* & \textbf{10.46}* & \textbf{10.44}* & \textbf{10.32}* & \textbf{15.49}* & 14.17* & \textbf{13.89}* & \textbf{13.75}* & \textbf{13.61}* \\
 \hline
 \hline
\multirow{9}{*}{\rotatebox{90}{LEAR}} & id & 15.38 & 14.84 & 15.22 & 16.57 & 16.07 & 11.00 & 10.41 & 10.11 & 10.20 & 10.18 & 14.78 & 14.32 & 14.48 & 14.66 & 14.37 \\
 & asinh & 14.49 & 14.12 & 14.22 & 14.87 & 13.97 & 10.75 & 10.41 & 10.20 & 10.16 & 10.06 & 14.41 & 13.96 & 13.88 & 13.83 & 13.18 \\
 & boxcox & 14.48 & 14.11 & 14.22 & 14.82 & {\ul 13.89} & 10.73 & 10.33 & 10.12 & {\ul 10.12} & {\ul 9.99} & 14.37 & 13.96 & 13.88 & 13.86 & {\ul 13.14} \\
 & mlog & 14.50 & 14.10 & 14.24 & 14.83 & 13.92 & 10.77 & 10.40 & 10.21 & 10.17 & 10.06 & 14.41 & 13.97 & 13.95 & 13.89 & 13.17 \\
 & T-PIT & {\ul 14.25} & {\ul 13.76} & {\ul 13.75} & {\ul 14.15} & 15.91 & {\ul \textbf{10.15}} & {\ul 9.93} & {\ul 9.88} & 10.20 & 11.47 & {\ul 14.04} & {\ul 13.61} & {\ul 13.44} & {\ul 13.63} & 15.08 \\[2pt]
 \cline{2-17}
 & $\mathrm{SL}_{\text{fix}}$ & 14.25 & 13.76 & 14.22 & 14.83 & 13.97 & \textbf{10.15} & 9.93 & 10.20 & 10.20 & 11.47 & 14.41 & 13.96 & 13.88 & 13.83 & 13.18 \\
 & $\mathrm{SL}_{\text{roll}}$ & 14.35 & 13.82 & 13.70 & 13.84* & 13.74* & 10.23 & 9.87* & 9.74* & 9.73* & 9.87* & 14.22 & 13.65 & 13.19* & 13.11* & 12.99* \\
 & $\mathrm{AVG}_{\text{fix}}$ & 14.25 & 13.76 & 14.22 & 14.83 & 13.90 & \textbf{10.15} & 9.93 & 10.20 & 10.20 & 11.47 & \textbf{13.84}* & \textbf{13.47}* & 13.88 & 13.83 & 13.18 \\
 & $\mathrm{AVG}_{\text{roll}}$ & \textbf{14.13}* & \textbf{13.71}* & \textbf{13.58}* & \textbf{13.63}* & \textbf{13.55}* & 10.15 & \textbf{9.84}* & \textbf{9.69}* & \textbf{9.67}* & \textbf{9.86}* & 14.03* & 13.54* & \textbf{13.11}* & \textbf{13.01}* & \textbf{12.97}*

\end{tabular}
}\\[3pt]
Note: * indicate that given model is significantly different from all individual transformation models at the 5\% level of the \cite{gia:whi:06} test.
\end{table*}

The variability in optimal parameter choices observed in Section 4.1 shows that fixed values cannot adequately capture the dynamics of different markets or calibration windows. This motivates the use of adaptive approaches that either select the best-performing transformation or combine forecasts from multiple VSTs. 

To address this in Table~\ref{tab:results} we report MAE values for individual variance-stabilizing transformations (upper subsections, averaged across parameter values $\mathrm{AVG}_{\text{roll}}(\cdot)$) and for the proposed selection and averaging schemes (lower subsections). Results are shown for NARX (upper panel) and LEAR (lower panel) across three markets and five calibration window lengths. Asterisks indicate cases where all five separate \citet{gia:whi:06} tests (each for one individual transformation) show statistical significance between models with and without selection or averaging procedure.  

The result strongly indicate that the rolling averaging scheme ($\mathrm{AVG}_{\text{roll}}$) consistently provides the strongest performance, reducing MAE even up to 17.7\% (with average 6.6\% across both model specifications, all markets and calibration window sizes) relative to benchmark without transformation (LEAR, EPEX-DE, 728-day window) and up to 14\% (with average 3.1\% across both model specifications, all markets and calibration window sizes) compare to model $\mathrm{SL}_{\text{roll}}$ with fixed selection transformation (LEAR, OMIE-ES, 1456-day window). Rolling selection ($\mathrm{SL}_{\text{roll}}$) typically follows as the next-best strategy, while fixed schemes lag behind. The statistical tests confirm these findings. The result of \citet{gia:whi:06} test shows that in most cases (28 out of 30 cases), $\mathrm{AVG}_{\text{roll}}$ delivers significantly better forecasts than any single transformation.

Among the individual transformations, the patterns differ substantially between the two model classes, but transformations remain very powerful tool to reduce forecast accuracy with improvement even up to 14.6\% for LEAR (EPEX-DE, 728-day window) and 8.7\% for NARX (EPEX-DE, 56-day window) compared to model without data transformation. 

For LEAR, T-PIT is the dominant choice, yielding the best single transformation results in 11 out of 15 cases across markets and windows. Box-Cox and asinh are occasionally competitive, while mlog never provides the lowest errors. On the other hand T-PIT is the worst choice (underperformed even compared to model without transformation for both LEAR and NARX schemes) for the longest 1456 day calibration window. It highlights that the transformation should not only be optimized for given market but also considering the size of calibration window size.

On the other hand, no single transformation clearly dominates the competitors for the NARX model. All transformations consistently outperform the model without this preprocessing step. However, accuracy gains decrease as the calibration window size increases.

\begin{figure*}[tb!]
 \includegraphics[width = 0.99\linewidth]{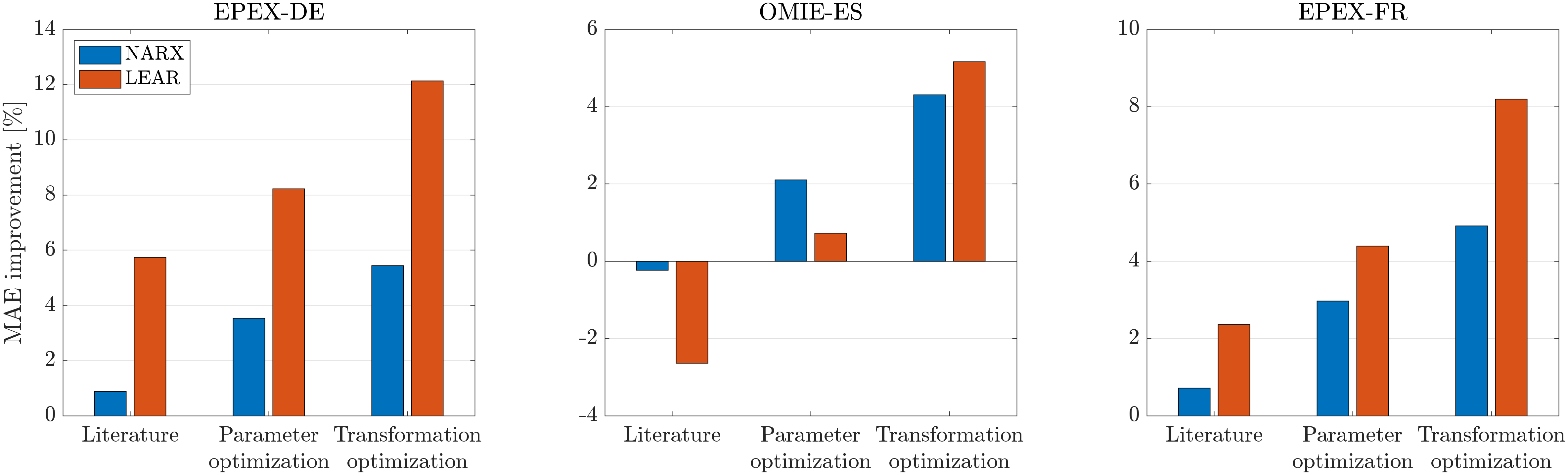}
 \caption{\rev{Average MAE improvements relative to the baseline model without variance-stabilizing transformation for three electricity markets and two base models (NARX and LEAR).}}
\label{fig:improvement}
\end{figure*}

\rev{To facilitate a direct comparison of the effects of adaptive parameterization and transformation selection, Figure~\ref{fig:improvement} reports relative improvements in forecasting accuracy with respect to the baseline model without variance-stabilizing transformation. 
For each configuration, the improvement is computed as
$
\Delta_{\mathrm{MAE}} = \frac{\text{MAE}_{\mathrm{id}} - \text{MAE}_i}{\text{MAE}_{\mathrm{id}}} \times 100\%,
$
where $\text{MAE}_{\mathrm{id}}$ denotes the MAE of the model estimated on untransformed prices and $\text{MAE}_i$ denotes the MAE obtained for a given variance-stabilizing transformation and selection scheme.

To be more specific the baseline level represents the average MAE of models estimated without variance-stabilizing transformation, averaged over all calibration window sizes. 
The bars labeled 'Literature' and 'Parameter optimization' report average MAE values obtained using variance-stabilizing transformations, averaged jointly over all considered transformations and calibration window sizes, with parameters selected either using fixed literature-based rules or adaptively using the rolling $\mathrm{AVG}_{\text{roll}}$ scheme, respectively. 
Finally, the bar labeled 'Transformation optimization' reports average MAE values averaged over all calibration window sizes for models in which the variance-stabilizing transformation itself is selected adaptively using the $\mathrm{AVG}_{\text{roll}}$ criterion.

The results confirm that static, literature-based parameter choices only yield limited improvements over the baseline, and in some cases are slightly worse. 
In contrast, adaptive parameter optimization leads to systematic gains across markets and base models, which increase further when the transformation itself is also selected adaptively. Overall, adaptive transformation selection provides the greatest and most consistent improvements in accuracy. 
These effects are more pronounced for the LEAR model than for NARX, indicating that the benefits of variance-stabilizing transformations are expected to weaken for more complex models and may diminish for highly flexible nonlinear structures such as deep neural networks \citep{lag:mar:des:wer:21} or hybrid and attention-based approaches \citep{ma:mei:22}.}

\subsection{Temporal Analysis of VST Performance}
\begin{figure*}[b!]
 \includegraphics[width = 0.99\linewidth]{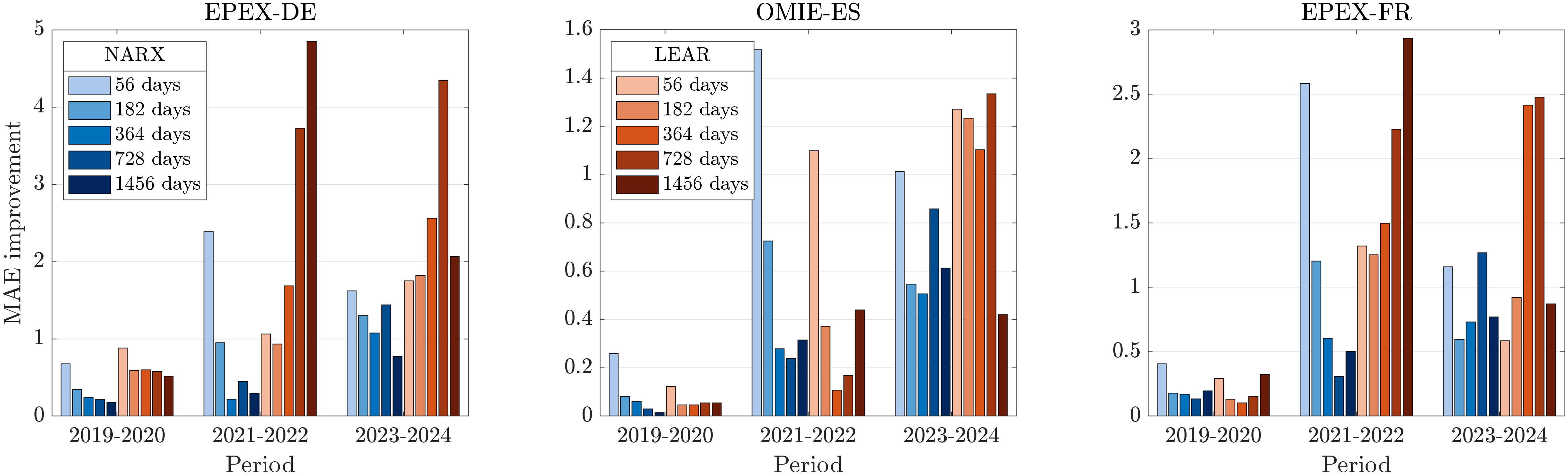}
 \caption{Improvements in forecast accuracy from variance stabilizing transformations relative to untransformed benchmarks, reported for the rolling averaging scheme ($\mathrm{AVG}_{\text{roll}}$). Results are shown for three subperiods (2019–2020, 2021–2022, 2023–2024), five calibration window sizes, and two model classes (NARX in blue, LEAR in orange).}
\label{fig:DeltaMAE}
\end{figure*}

To examine whether the effectiveness of variance stabilizing transformations depends on current market conditions, the out-of-sample period (21.02.2019–31.12.2024) is divided into three phases: 2019–2020 (pre-shock), 2021–2022 (COVID recovery and onset of the energy crisis), and 2023–2024 (high-volatility regime). 

Figure~\ref{fig:DeltaMAE} summarizes the results for the $\mathrm{AVG}_{\text{roll}}$ model, where for each market (EPEX-DE, OMIE-ES, EPEX-FR) the improvement in MAE over the no-transformation benchmark is shown across calibration window sizes. Blue bars correspond to NARX models and orange bars to LEAR models, with color intensity increasing from shorter to longer windows.

In the presented result we can observe several patterns. First, the benefits of VSTs are strongly regime-dependent: across all markets, gains are modest in 2019–2020 but increase substantially in 2021–2022 and remain elevated in 2023–2024. This suggests that transformations become particularly valuable during volatile periods.

Second, the two model classes respond differently to window length. For NARX, shorter calibration windows yield the largest gains. In contrast, LEAR generally benefits from medium to long windows. 

Third, the magnitude of improvements differs substantially across markets. Germany shows the strongest gains, with MAE reductions up to 5 Euro, particularly in 2021–2022 and 2023–2024. France also benefits considerably, with improvements reaching 3 Euro, especially in the crisis period of 2021–2022. By contrast, Spain exhibits much smaller gains, rarely exceeding 1.5 Euro, suggesting that VSTs play a less critical role in stabilizing its price dynamics.

\section{Conclusions}
\label{sec:conclusion}

This study examined the role of variance stabilizing transformations (VSTs) in day-ahead electricity price forecasting across three liberalized markets -- Germany, Spain, and France --over the period 2015–2024. By analyzing four widely used transformations (asinh, Box–Cox, mirror logarithm, and T-PIT) within two distinct model classes (NARX and LEAR), we show that the choice of transformation and its parameterization critically influences forecast accuracy, with no single transformation performs best across all markets, horizons, and calibration windows

The empirical evidence demonstrates that VSTs substantially enhance forecast performance. The best-performing individual transformation improves MAE by up to 14.6\% for LEAR and 8.7\% for NARX relative to untransformed benchmarks. These benefits are particularly pronounced during periods of increased volatility, underscoring the effectiveness of VSTs in stabilizing dynamics when price fluctuations are most severe. A further contribution is the introduction of a novel parametrization of the asinh transformation, which consistently outperforms its standard form and yields significant accuracy gains across all markets and models.

To overcome the absence of a universally optimal transformation, we proposed adaptive schemes that either select or average forecasts across multiple parameterizations. Among them, the rolling averaging scheme ($\mathrm{AVG}_{\text{roll}}$) delivers the strongest and most consistent gains, reducing forecast errors by up to 17.7\% compared to untransformed models and outperforming fixed or single-transformation approaches in nearly all settings.

Overall, the findings establish VSTs—particularly when combined with adaptive averaging—as a powerful tool to improve electricity price forecasting under volatile market conditions. Limitations include the focus on point forecasts and a restricted set of transformations. Future work may extend the analysis to probabilistic forecasting and explore additional transformation families.

\section*{Acknowledgements}
The study was partially supported by the National Science Center (NCN, Poland) through grant no.\ 2021/43/I/HS4/02578

\section*{Declaration of generative AI}
During the preparation of this work the author used ChatGPT (Open AI) in order to language editing. After using this tool/service, the author(s) reviewed and edited the content as needed and take(s) full responsibility for the content of the published article.

\setstretch{1.0}
\bibliographystyle{elsarticle-harv}
\bibliography{Uniejewski_VST}

\end{document}